\documentclass[12pt]{article}
\usepackage{amssymb,graphicx}

\begin{document}

\title{Superbranes and Generic Curved Spacetime}
\author{\bf{Dj. \v{S}ija\v{c}ki}
\thanks{\,e-mail address: sijacki@phy.bg.ac.yu} \\
\normalsize{Institute of Physics P.O.Box 57, 11001 Belgrade, Serbia}}

\date{}
\maketitle

\begin{abstract}
Embedding of a bosonic and/or fermionic p-brane into a generic curved
$D$-dimensional spacetime is considered. In contradistinction to the bosonic
p-brane case,  when there are no constraints on a generic curving whatsoever,
the usual superbrane can be embedded into a curved spacetime of a restricted
curving only. A generic curving is achieved by extending the odd sector of a
superbrane as to transform w.r.t. $\overline{SL}(D,R)$,
i.e. $\overline{Diff}(D,R)$ infinite-component spinorial
representations. Relevant constructions in the $D=3$ case are considered.
\end{abstract}

\section{Generic curved target spacetime for \\ superbrane}

In the conventional lagrangian formulation for superbranes, the
$(p+1)$-dimensional curved (locally reparametrizable) brane world
sheet/volume $R^{p+1}$ is embedded in a flat (Poincar\'e invariance) Minkowski
space-time $M^{1,D-1}$.

On the other hand, macroscopic gravity is described classically by
Einstein's theory, corresponding to a generic curved Riemannian $R^4$
manifold (general covariance).

Thus one is faced with an apparent difference in the manifest
symmetries of these two theories. This difference is not only of the
principal nature, but is crucial for numerous practical questions such
as nonperturbative gravitational solutions (Schwarzshild) etc.

One can certainly hope to reconstruct the full general covariance
starting from the field theory of superbrane embedded in a flat
space. However, preliminary difficulties encountered along this line
support a more pragmatic (and in our opinion in fact the only)
approach to construct an a priori fully generally-covariant target-space
superbrane theory.

\subsection{Bosonic brane: Flat to curved space}

The (bosonic) p-brane action [1],
\begin{eqnarray*}
&S& = \int d^{p+1}\xi \ \bigg( \frac{1}{2} \sqrt{-\gamma}
 \gamma^{ij}(\xi) \partial_i X^m \partial_j X^n \eta_{mn}
 -\frac{1}{2}(p-1)\sqrt{-\gamma} \\
&&+ \frac{1}{(p+1)!}
 \epsilon^{i_{1}i_{2}\cdots i_{p+1}}  \partial_{i_{1}}X^{m_{1}}
 \partial_{i_{2}}X^{m_{2}} \cdots  \partial_{i_{p+1}}X^{m_{p+1}}
 A_{m_{1}m_{2}\cdots m_{p+1}}(X) \bigg) ,
\end{eqnarray*}
where $i$ $=$ $0,1,\dots p$ labels the coordinates $\xi^i =
(\tau,\sigma,\rho,\dots)$ of the brane world-volume with metric
$\gamma_{ij}$, and  $\gamma=\det(\gamma_{ij})$; $m$ $=$ $0,1,\dots
,D-1$ labels the target-space coordinates $X^{m}$ with metric
$\eta_{mn}$, and $A_{m_{1}m_{2}\dots m_{p+1}}$ is a ($p + 1$)-form
characterizing a Wess-Zumino-like term {\em can be generalized} in a
straight forward way {\em for a generic curved target space} to read
\begin{eqnarray*}
&S& = \int d^{p+1}\xi \
\bigg(\frac{1}{2}\sqrt{-\gamma}\gamma^{ij}(\xi)
\partial_iX^{\tilde{m}} \partial_jX^{\tilde{n}}
g_{\tilde{m}\tilde{n}} -{1\over2}(p-1)\sqrt{-\gamma} \\ 
&&+ \frac{1}{(p+1)!} \epsilon^{i_{1}i_{2}\cdots i_{p}+1}
\partial_{i_{1}}X^{\tilde{m}_{1}}
\partial_{i_{2}}X^{\tilde{m}_{2}}\cdots
\partial_{i_{p+1}}X^{\tilde{m}_{p+1}}
A_{\tilde{m}_{1}\tilde{m}_{2}\cdots \tilde{m}_{p+1}}(X) \bigg) ,
\end{eqnarray*}
where $\tilde{m}$ $=$ $0,1,\dots ,D-1$ labels the curved target-space
coordinates $X^{\tilde{m}}$ with riemannian metric
$g_{\tilde{m}\tilde{n}}$

\begin{center}
\begin{tabular}{ccc}
 $SO(1,D-1):\quad$ & $X^m$        & $\eta_{mn}$ \\ $\downarrow$ &
 $\downarrow$     & $\downarrow$ \\ $Diff(D,R):$      &
 $X^{\tilde{m}}$  & $g_{\tilde{m}\tilde{n}}$ \\
\end{tabular}
\end{center}

\subsection{Super brane: Flat to curved space}

The super p-brane action reads [2]:
\begin{eqnarray*}
&S& = \int d^{p+1} \xi \bigg( \frac{1}{2}
\sqrt{\gamma}\gamma^{ij}(\xi) \Pi^m_{i} \Pi^n_{j} \eta_{mn}
-\frac{1}{2}(p-1) \sqrt{-\gamma} \\
&&+ \frac{1}{(p+1)!}
\epsilon^{i_{1}i_{2}\cdots i_{p+1}}
\partial_{i_{1}}Z^{a_{1}}\partial_{i_{2}}Z^{a_{2}}\cdots
\partial_{i_{p+1}}Z^{a_{p+1}}  B_{a_{p+1}\cdots a_{2}a_{1}} \bigg)\ .
\end{eqnarray*}
Here, the target space is a supermanifold with super-space coordinates
$Z^{a}=(X^{m},\Theta^{\alpha})$, $\Pi^{m}_{i} = \partial_{i}X^{m} -
\bar\Theta\Gamma^{m}\partial_{i}\Theta$, where $m=0,1,\cdots ,D-1$,
$\alpha = 1,2,\cdots , 2^{\left[\frac{D-1}{2}\right]}$, and
$\Gamma^{m}$ are the corresponding $D$-dimensional spacetime gamma
matrices.

Note that $\Theta^{\alpha}$ transforms w.r.t. fundamental spinorial
representation of the $Spin(1,D-1)$ $\simeq$ $\overline{SO}(1,D-1)$ group.

In contradistinction to the bosonic brane case where, while spacetime
curving, the $SO(1,D-1)$ group was replaced by the $Diff(D,R)$ one,
here in the super brane case, the $Spin(1,D-1)$ group is to be
replaced by the covering group of the General Coordinate
Transformations group $GCT$, i.e. $\overline{Diff}(D,R)$.

There are no finite-dimensional representations of the $\overline{Diff}(D,R)$
group for $D \geq 3$ (cf. [3]), and thus {\em one cannot proceed as in the
bosonic case by systematically replacing all local (flat-space) tensorial
quantities by the appropriate world (curved-space) ones}.

\section{Topology and dimensionality of the \\ 
$\overline{Diff}(D,R)$ groups}

Topology of the $Diff(D,R)$  group, as well as of its $GL(D,R)$ and $SL(D,R)$
linear subgroups, is determined by the topology of its maximal compact
subgroup $SO(D)$, which is for $D\geq 3$ double connected ($G = KAN$;
the Abelian $A$ and nilpotent $N$ subgroups are contractible to a point and
therefore irrelevant for the topology questions). For pin/spin discussion
cf. [4].

Thus, in the quantum case, all these groups, for $D \geq 3$, have double
valued spinorial representations besides the usual tensorial ones.

\subsection{$Diff(D,R)$, $SL(D,R)$ covering groups}

The group-subgroup relations of the relevant groups for our
considerations is as follows:
\begin{center}
\begin{tabular}{ccccccc}
 1 &   & 1 &   & 1 &    & 1 \\ $\downarrow$ &  & $\downarrow$ &  &
 $\downarrow$ &  & $\downarrow$ \\ $Z_{2}$ &   &  $Z_{2}$ &   &
 $Z_{2}$ & &  $Z_{2}$ \\ $\downarrow$ &  & $\downarrow$ &  &
 $\downarrow$ &  & $\downarrow$ \\ $\overline{Diff}(D,R)$ & $\supset$
 & $\overline{GL}(D,R)$ & $\supset$ & $\overline{SL}(D,R)$ & $\supset$
 & $Spin(1,D-1)$ \\  $\downarrow$ &  & $\downarrow$ &  & $\downarrow$
 &  & $\downarrow$ \\ $Diff(D,R)$ & $\supset$ & $GL(D,R)$ & $\supset$
 & $SL(D,R)$ & $\supset$ & $SO(1,D-1)$ \\  $\downarrow$ &  &
 $\downarrow$ &  & $\downarrow$ &  & $\downarrow$ \\ 1 &   & 1 &   & 1
 &    & 1 \\
\end{tabular}
\end{center}

\subsection{$\overline{Diff}(D,R)$, $\overline{SL}(D,R)$ groups of matrices}

It turns out that {\em there are no finite-dimensional complex matrix groups}
that contain the $SL(D,R)$ $\supset$ $SO(D)$, $D\geq 3$ group-chain as
subgroups [3,5]. Moreover, $\overline{SL}(D,R)$, $D \geq 3$, the double
covering of $SL(D,R)$, is a group of infinite matrices. Thus, {\em all
spinorial representations of the $\overline{Diff}(D,R)$, $\overline{GL}(D,R)$,
$\overline{SL}(D,R)$ groups, for $D \geq 3$ are infinite-dimensional}, and
when restricted to the spacial $Spin(D-1)$ subgroup they contain all spins.

For example (cf. [6]), the simplest spinorial $\overline{SL}(3,R)$
representation from the (Ladder) Degenerate Series
$D^{ladd}_{\overline{SL}(3,R)}(\frac{1}{2})$ contains the following
$Spin(3)$ $\simeq$ $SU(2)$ representations:
$$ D^{\frac{1}{2}},\quad D^{\frac{5}{2}},\quad D^{\frac{9}{2}},\quad
etc.,
$$  while the representation $D^{pr}_{\overline{SL}(3,R)}(\frac{1}{2},
\sigma_{2}, \delta_{2})$ from the Principal Series contains:
$$ D^{\frac{1}{2}},\quad  2\times D^{\frac{3}{2}},\quad 3\times
D^{\frac{5}{2}},\quad etc.
$$

\section{Generic curved target-spacetime embedding}

In the standard approach to GR, spinors are defined w.r.t. a local
tangent spacetime and transform w.r.t. the local Lorentz symmetry group
$Spin(1,D-1)$, i.e. $SL(2,C)$ $\simeq$ $Spin(1,3)$ for $D=4$. The curved
spacetime (coordinates $x^{\mu}$) and the local Minkowskian one (coordinates
$x^{m}$)are mutually connected by the frame fields $e^a_{\mu}(x)$ (tetrads for
$D=4$). Analogous situation persists in the metric-affine [7] and/or
gauge-affine [8] case as well.

In the p-brane case, $Z^{a}=(X^{m},\Theta^{\alpha})$ defines a flat
tangent superspace over a curved p-brane spacetime at $\xi^{i}$.

In a parallel to GR, spinors of a curved spacetime of coordinates
$X^{m}$ are to be defined w.r.t. a "new" tangent spacetime erected at
every point $X^{m}$. In other words, {\it in order to define curved
target-space spinors one has to construct a flat tangent space to the
bosonic spacetime sector of a superbrane at every point $\xi$},
i.e. to a space that is itself a tangent space. Such a construction
simply does not exists, therefore {\em one can not define spinors of a
superbrane in a generic curved spacetime in the standard manner}
[9,10]. However, superbranes can be defined (in the standard way) for special
spacetimes (e.g. De Sitter, anti De Sitter, ...).

\subsection{Restricted curving}

Restricted curving is achieved by staying with finite tangent space \\
$Spin(1,D-1)$ spinors, but restricting further curving of M$^{1,D-1/r\cdot N}$
to such as can be described by that "diagonal" 
subgroup of $\overline{Diff}(M^{1,D-1/r\cdot N})$ that preserves the
orbits of $Spin(1,D-1)$ when acting simultaneously on both even and
odd sectors of superspace.  In other words, {\em allow no linear
transformations other than} $Spin(1,D-1)$ and adjoin a restricted set
of non-linear ones leading to manifolds carrying the action of
$Spin(1,D-1)$.

This method inherited from supergravity, has been used extensively in
the attempts to curve the "target space" in superstrings and in
supermembranes. It allows the highly  restricted rheonomic curving
undergone  by superspace in supergravity in which the group parameters
are constrained so that the odd coordinates are not gauged over.

The supertranslations act anholonomically as Lie derivatives
("anholonomized" general coordinate transformations), i.e. as part of
the curved-space modified structure group acting as an effective fibre
in the appropriate principle bundle.

The superbrane action for the {\em restricted curving} reads:

\begin{eqnarray*}
&S& = \int d^{p+1}\xi \bigg(  \frac{1}{2} \sqrt{\gamma}
\gamma^{ij}(\xi) E^{\tilde{a}}_i E^{\tilde{b}}_j
g_{\tilde{m}\tilde{n}} \\ 
&&- \frac{1}{2}(p-1) \sqrt{-\gamma}  +
\frac{1}{(p+1)!} \epsilon^{i_{1}i_{2}\cdots i_{p+1}}
E^{\tilde{a}_{1}}_{i_{1}}E^{\tilde{a}_{2}}_{i_{2}}\cdots
E^{\tilde{a}_{p+1}}_{i_{p+1}} B_{\tilde{a}_{p+1}\cdots
\tilde{a}_{2}\tilde{a}_{1}} \bigg)\ .
\end{eqnarray*}

Here, the target space is a supermanifold with super-space coordinates
$Z^{\tilde{a}}=(X^{\tilde{m}},\Theta^{\alpha})$, where $\tilde{m} =
0,1,\dots ,D-1$ and $\alpha = 1,2,\dots , 2^{\left[ \frac{D-1}{2}
\right]}$. Furthermore, $E^{a}_i = (\partial_i Z^{\tilde{a}})
E^{a}_{\tilde{a}}(Z)$, where $E^{a}_{\tilde{a}}$ is the supervielbein
and $a = (m\ \alpha)$ is the tangent-space index. In the standard
superspace formalism one tends to describe $\Theta^{\alpha}$ as a
"world" fermionic coordinate, but this time in a very {\em restricted
sense} only.

\subsection{Non-linear curving}

It is possible to use finite $Spin(1,D-1)$ spinors and represent  the quotient
$\overline{Diff}(D,R)$/$Spin(1,D-1)$ non-linearly over the $Spin(1,D-1)$
subgroup, following the pioneering work of Ogievetski and Polubarinov
[11]. The result is effectively that of the restricted curving. 

In the core of the corresponding non-linear representations is the non-linear
realizer field (the metric) 
$$ 
g_{\tilde{m}\tilde{n}} = \eta_{mn} e^{m}_{\tilde{m}}
e^{n}_{\tilde{n}}
$$
that defines the linear--to--nonlinear transformation:
\begin{eqnarray*}
&&L(g_{\tilde{m}\tilde{n}}) = exp(i g_{\tilde{m}\tilde{n}}
    T^{\tilde{m}\tilde{n}} ) , \quad\ \ T^{\tilde{m}\tilde{n}} \in
    sl(D,R)/spin(1,D-1) ,\\ 
&&\overline{Diff}(D,R)/Spin(1,D-1) \\ 
&&\quad\quad\quad = \overline{Diff}(D,R)/\overline{SL}(D,R)\times
    \overline{SL}(D,R)/Spin(1,D-1). \\
\end{eqnarray*}
Mathematical consistency of a curved superspace, i.e. a mutual relation of the
bosonic sector given by non-linear curving and the fermionic sector given by
$Spin(1,D-1)$ representations, imposes constraints equivalent to those of the
restricted curving.

\subsection{Generic curving}

In the generic curving case we make use of (infinite) world spinors
transforming w.r.t. the covering group of the General Coordinate
Transformations, $\overline{GCT} = \overline{Diff}(D,R)$. This approach for
the superstring was initiated together with Yuval Ne'eman [9]. There are two
possible scenarios:

1. {\it "Minimal" solution -- change in the fermionic sector only}:\\
Here, we replace
$$ 
\Theta^{\alpha}, \quad \alpha = 1,\dots ,  2^{\left[ \frac{D-1}{2}
\right]} ;\quad \quad   \Theta \sim Rep(Spin(D))
$$  
by a corresponding world spinor
$$ 
{}\quad\quad \Theta^{\tilde{A}}, \quad \tilde{A} = \frac{1}{2},
\dots ,\infty ;\quad \quad \quad  \Theta \sim
Rep(\overline{Diff}(D,R)).
$$

2. {\it "Maximal" solution -- "world" superspace formulation (generic curved
superspace supersymmetry)}:\\  
Here we replace
$$ 
Z^{a} = (X^{m}, \Theta^{\alpha}) ;\quad\quad X,\ \Theta \sim
Rep(Spin(D))  \quad\quad
$$ 
by a corresponding curved superspace coordinates
$$ 
Z^{\tilde{I}} = (X^{\tilde{M}}, \Theta^{\tilde{A}}) ;\quad\quad X,\
\Theta \sim Rep(\overline{Diff}(D,R)) ,
$$ 
that are of infinite range for both bosonic and fermionic coordinates. The
appropriate replacements are as follows: 

\begin{center}
\begin{tabular}{ccccccccc}
$Spin(1,D-1):\quad$ & $X^m$ & $\eta_{mn}$ & $\quad$ &
$\Theta^{\alpha}$ & $\gamma^{m}$ & $\quad$ & $X^m$ & $\eta_{mn}$ \\  
$\downarrow$ & $\downarrow$ & $\downarrow$ & $\quad$ &
$\downarrow$ & $\downarrow$ & $\quad$ & $\downarrow$ &
$\downarrow$ \\  
$\overline{SL}(D,R):\quad$ & $X^m$ & $\eta_{mn}$ & $\quad$
& $\Theta^{A}$ & $\Gamma^{m}_{(SL)}$ & $\quad$ & $X^{M}$ &
$\eta_{MN}$ \\ 
$\downarrow$ & $\downarrow$ & $\downarrow$ &
$\quad$ & $\downarrow$ & $\downarrow$ & $\quad$ &
$\downarrow$ & $\downarrow$\\  
$\overline{Diff}(D,R):$ & $X^{\tilde{m}}$ &
$g_{\tilde{m}\tilde{n}}$ & $\quad$ & $\Theta^{\tilde{A}}$ &
$\Gamma^{\tilde{m}}_{(Diff)}$ & $\quad$ & $X^{\tilde{M}}$ &
$G_{\tilde{M}\tilde{N}}$ \\
\end{tabular}
\end{center}

\section{Group-theoretical constructions for a generic curved
  spacetime superbrane embedding}

\noindent {\bf (i)} Spinorial and infinite-dimensional tensorial
representations of the \\ $\overline{SL}(D,R)$ group. 

\noindent {\bf (ii)}  Spinorial and infinite-dimensional tensorial
representations of the \\ $\overline{Diff}(D,R)$ group

\noindent {\bf (iii)} Dirac-like equation for $\overline{SL}(D,R)$ and
$\overline{Diff}(D,R)$ spinors, i.e. the corresponding (infinite) 
$\Gamma^{m}_{(SL)}$, $\Gamma^{\tilde{m}}_{(Diff)}$ 
generalizations of the $\gamma$ matrices, which a required for the
expressions such as:
$$ 
E_{i}^{m}\ \rightarrow\ E_{i}^{\tilde m} = \partial _{i} X^{\tilde m} -
i\overline{\Theta }^{\tilde A} 
(\Gamma ^{\tilde{m}}_{(Diff)})_{\tilde A\tilde B} 
\partial _{i}\Theta^{\tilde B},
$$

\noindent {\bf (iv)} Infinite super algebras that generalize the
Virasoro and Neveu-Schwarz-Ramond ones and contain respectively the
$SL(D,R)$ and $\overline{SL}(D,R)$   tensorial and spinorial adjoint
representations as subalgebras, thus providing for a complete
superspace supersymmetry formulation.

\section{$\overline{SL}(D,R)$ Spinorial representations - $Spin(D)$
multiplicity free case}

The $\overline{SL}(D,R)$ group can be contracted (a la
Wigner-In\" on\" u) w.r.t. its $\overline{SO}(D)$ subgroup to yield
the semidirect-product group $T'\wedge \overline{SO}(D)$. $T'$ is an
Abelian group generated by operators $U_{mn}$, which form an
$\overline{SO}(D)$ second rank symmetric operator with commutation
relations
$$ [J,\ J] \subset J, \quad\quad [J,\ U] \subset U, \quad\quad [U,\ U]
= 0.
$$  An efficient way of constructing explicitly the
$\overline{SL}(D,R)$ infinite-dimensional representations is based on
the decontraction formula, which is an inverse of the Wigner-In\" on\"
u contraction.  According to the decontraction formula, the following
operators [12]
$$ T_{mn} = pU_{mn} + {i\over{2\sqrt{U\cdot U}}} \left[
 C_2(\overline{SO}(D)),\ U_{mn}\right],
$$  together with $J_{mn}$ form the $\overline{SL}(D,R)$ algebra. The
parameter $p$ is an arbitrary complex number, and
$C_2(\overline{SO}(D))$ is the $\overline{SO}(D)$ second-rank Casimir
operator.
 
For the representation Hilbert space we take the homogeneous space of
$L^2$ functions of the maximal compact subgroup $\overline{SO}(D)$
parameters. The $\overline{SO}(D)$ representation labels are given
either by the Dynkin labels $({\lambda}_1, {\lambda}_2,\dots ,
{\lambda}_r)$ or by the highest weight vector which we denote by $\{
j\} = \{ j_1, j_2, \dots , j_r \}$, $r = \left[ D\over 2\right]$.

The $\overline{SL}(D,R)$ commutation relations are invariant w.r.t. an
automorphism defined by:
$$
s(J) = +J, \quad\quad s(T) = -T .
$$ 
This enables us to
define an '$s$-parity' to each $\overline{SO}(D)$ representation of an
$\overline{SL}(D,R)$ representation. In terms of Dynkin labels we find
\begin{eqnarray*}
s(D_2) &=& (-)^{{1 \over 2}({\lambda}_1 + {\lambda}_2 - \epsilon )},
\\ s(D_{n\ge 3}) &=& (-)^{ {\lambda}_1 + {\lambda}_2 + \dots +
{\lambda}_{n-2} + {1\over 2}({\lambda}_n - {\lambda}_{n-1} - \epsilon
)} \\ s(B_1) &=& (-)^{ {1\over 2}({\lambda}_1 - \epsilon )} \\
s(B_{n\ge 2}) &=& (-)^{ {\lambda}_1 + {\lambda}_2 + \dots +
{\lambda}_{n-1} + {1\over 2}({\lambda}_n - \epsilon )} \\
\end{eqnarray*}
where $\epsilon = 0\ (+1)$ if $\lambda$ is even (odd).

For the ${1\over 2}(D+2)(D-1)$-dimension representation of
$\overline{SO}(D)$, i.e. for $(20\dots 0) = \Box\!\Box $, one has $s(20\dots
0) = +1$. A basis of an $\overline{SO}(D)$ representation is provided by
the Gel'fand - Zetlin pattern characterized by the maximal weight
vectors of the subgroup chain $\overline{SO}(D) \supset
\overline{SO}(D-1) \supset \overline{SO}(2)$. We write the basic
vectors as $\left| { \{ j \} }\atop { \{m \} }\right>$, where $\{ m \}$
corresponds to $\overline{SO}(D-1) \supset \overline{SO}(D-2) 
\supset \overline{SO}(2)$ subgroup chain weight vectors.

The Abelian group generators $\{ U\} = U_{\{ \mu \} }^{[\Box\!\Box \}
}$ can be, in the case of multiplicity free representations, written
in terms of the $\overline{SO}(D)$-Wigner functions as follows $U_{ \{
\mu \} }^{\{ \Box\!\Box \} } = D_{ \{ 0\} \{ \mu \} }^{\{ \Box\!\Box
\} }(\phi )$. It is  now rather straightforward to determine the
noncompact operators matrix elements, which read [5,12]
\begin{eqnarray*}
\left< \begin{array}{c} \{ j'\} \\ \{ m'\} \end{array} \right|  
T_{\{\mu \}}^{\{ \Box\!\Box \}}  \left| 
\begin{array}{c} \{ j\} \\ \{ m\} \end{array} \right>  
= \left( \begin{array}{ccc} \{ j'\} & \{\Box\!\Box \} & \{ j\} \\ 
\{ m'\} & \{ \mu \} & \{ m\} \end{array} \right)  
< \{ j'\} || T^{\{ \Box\!\Box \} } || \{ j \} >, 
\\ 
< \{ j'\} || T^{\{ \Box\!\Box \} } || \{ j\} > = 
\sqrt{dim\{ j'\} dim\{ j\} }  \left\{ p + 
\frac{1}{2} ( C_2 (\{ j'\}) - C_2 (\{ j\} ) ) \right\} 
\\
\times \left(
\begin{array}{ccc} \{ j'\} & \{ \Box\!\Box \} & \{ j\} \\ 
\{ 0\} & \{ 0\} & \{ 0\} \end{array}   \right). \\
\end{eqnarray*}
\noindent $\left( \begin{array}{ccc} \cdot & \cdot & \cdot \\ \cdot & \cdot&
\cdot \end{array} \right) $  is the appropriate $"3j"$ symbol for the
$\overline{SO}(D)$ group. For the multiplicity free $\overline{SL}(D,R)$
representations each $\overline{SO}(D)$ sub-representation appears at most
once and has the same $s$-parity.

\section{$\overline{Diff}(D,R)$ representations for world spinor fields}

The world spinor fields transform w.r.t. $\overline{Diff}(D,R)$ as follows
\begin{eqnarray*}
&&(D(a,\bar f)\Psi_M) (x) = (U_{ \overline{Diff}_0(D,R)}(\bar f))^N_M
\Psi_N (f^{-1}(x-a)), \\ &&(a,\bar f) \in T_{D} \wedge
\overline{Diff}_0(D,R),  \nonumber
\end{eqnarray*}
\noindent where $\overline{Diff}_0(D,R)$ is the homogeneous part
of $\overline{Diff}(D,R)$, while $f$ is the element corresponding to
$\bar f$ in $Diff(D,R)$. The $D_{\overline{Diff}_0(D,R)}$
representations can be reduced to direct sum of infinite-dimensional
$\overline{SL}(D,R)$ representations. We consider here those
representations of $\overline{Diff}_0(D,R)$ that are nonlinearly
realized over the maximal linear subgroup $\overline{SL}(D,R)$.

Provided the relevant $\overline{SL}(D,R)$ representations are known,
one can first define the corresponding general/special affine spinor
fields, $\Psi_{A}(x)$, and than make use of the infinite-component
pseudo-frame fields $E^{A}_{\tilde{A}}(x)$ (linear-to-nonlinear
mapping) [13,14],
$$ \Psi_{\tilde{A}}(x) = E^{A}_{\tilde{A}}(x) \Psi_{A}(x), \quad\quad
E^{A}_{\tilde{A}}(x) \sim \overline{Diff}_{0}(D,R)/\overline{SL}(D,R)
$$ where $\Psi_{\tilde{A}}(x)$ and $\Psi_{A}(x)$ are the world
(curved-space) and local Affine (flat-space) spinor fields
respectively.

Their infinitesimal transformations are
$$ \delta E^{A}_{\tilde{A}}(x) = i\epsilon^{a}_{b}(x)
\{Q_{a}^{b}\}^{A}_{B} E^{B}_{\tilde{A}}(x) + \partial_\mu \xi^{\nu}
e^{a}_{\nu} e^{\mu}_{b}\{Q^{a}_{b}\}^{A}_{B} E^{B}_{\tilde{A}}(x),
$$ where  $\epsilon^{a}_{b}$ and $\xi^\mu$ are group parameters of
$\overline{SL}(D,R)$ and \\
$\overline{Diff}(D,R)/\overline{Diff}_{0}(D,R)$ respectively, while
$e^{a}_{\nu}$ are the standard $n$-bine frame fields.

The transformation properties of the world spinor fields themselves
are given as follows:
$$ \delta \Psi^{\tilde{A}}(x) = i\big\{ \epsilon^{a}_{b}(x)
E^{\tilde{A}}_{A}(x) \big(Q_{a}^{b}\big)^{A}_{B} E^{B}_{\tilde{B}}(x)
+ \xi^{\mu} \big[ \delta^{\tilde{A}}_{\tilde{B}}\partial_{\mu} +
E^{\tilde{A}}_{B}(x)\partial_{\mu}E^{B}_{\tilde{B}}(x)\big] \big\}
\Psi^{\tilde{B}}(x).
$$ The $\big( Q^{b}_{a} \big)^{\tilde{A}}_{\tilde{B}} =
E^{\tilde{A}}_{A}(x) \big(Q_{a}^{b}\big)^{A}_{B} E^{B}_{\tilde{B}}(x)$
is the holonomic form of the $\overline{SL}(D,R)$ generators given in
terms of the corresponding anholonomic ones. The $\big( Q^{b}_{a}
\big)^{\tilde{A}}_{\tilde{B}}$  and  $\big(Q_{a}^{b}\big)^{A}_{B}$ act
in the spaces of spinor fields $\Psi_{\tilde{A}}(x)$ and $\Psi_{A}(x)$
respectively.

The above outlined construction allows one to define a fully
$\overline{Diff}(D,R)$ covariant Dirac-like wave equation for the
corresponding world spinor fields provided a Dirac-like wave equation
for the $\overline{SL}(D,R)$ group is known. In other words, one can
lift up an $\overline{SL}(D,R)$ covariant equation of the form
$$ \big( ie^{\tilde{m}}_{m} \big(\Gamma^{m}_{(SL)}\big)^{B}_{A}
\partial_{\tilde{m}} - \mu \big) \Psi_{B}(x) = 0,  \nonumber
$$ to a $\overline{Diff}(n,R)$ covariant equation
$$ \big( ie^{\tilde{m}}_{m} E^{A}_{\tilde{A}}
\big(\Gamma^{m}_{(SL)}\big)^{B}_{A}E^{\tilde{B}}_{B}  \partial_{m} -
\mu \big) \Psi_{\tilde{B}}(x) = 0,
$$ where the former equation exists provided a spinorial
$\overline{SL}(D,R)$ representation for $\Psi$ is given, such that the
corresponding representation Hilbert space is invariant
w.r.t. $\Gamma^{m}_{(SL)}$  action. Thus, the crucial step towards a
Dirac-like world spinor equation is a construction of the vector
operator $\Gamma^{m}_{(SL)}$ in the space of $\overline{SL}(D,R)$
spinorial representations [5,15].

\section{$\Gamma^{m}_{(SL)}$ for a Dirac-like world spinor equation}

It is well known that one can satisfy the commutation relations
$$ [M_{mn}, \Gamma_{p}] = i(\eta_{mp}\Gamma_n - \eta_{np}\Gamma_m),
\quad\quad M_{mn} \in spin(1,D-1) ,
$$ in the Hilbert space of $Spin(1,D-1)$ irreducible representations.
However, in order for an $Spin(1,D-1)$ vector to be an
$\overline{SL}(D,R)$ vector as well, it has to satisfy additionally
the following commutation relations
$$ [T_{mn}, \Gamma_{p}] = i(\eta_{mp}\Gamma_n + \eta_{np}\Gamma_m) ,
\quad\quad T_{mn} \in sl(D,R)/spin(1,D-1) .
$$ This is a much harder task to achieve [16], and in principle, one can
find nontrivial solutions only for particular representation spaces.

Example: For $SL(3,R)$ finite-dimensional reps., one can satisfy the
above algebraic conditions only in the special case of a reducible
representation of Young tableaux $[2q+1,q] \oplus [2q+1,q+1]$.

The multiplicity free (ladder) unitary (infinite-dimensional)
irreducible representations

$$ D^{ladd)}_{SL(3,R)}(0,\sigma_2 ), \quad\quad \{ j\} = \{ 0,2,4,
\dots \} ,
$$  and
$$ D^{ladd)}_{SL(3,R)}(1,\sigma_2 ), \quad\quad \{ j\} = \{ 1,3,5,
\dots \},
$$ can be viewed as limiting cases of the series of finite-dimensional
representations $[0,0]$, $[2,0]$, $[4,0]$, ..., and $[1,0]$, $[3,0]$,
$[5,0]$, ... respectively.

Upon the coupling with the $SL(3,R)$ vector representation $[1,0]$,
one has $[1,0]\otimes [2n,0] \supset [2n+1,0]$, and $[1,0]\otimes
[2n+1,0] \supset [2n+2,0]$, ($n=0,1,2,\dots$). It seems possible to
represent the vector operator $\Gamma^{m}$ in the Hilbert space of the
$D^{ladd)}_{SL(3,R)}(0,\sigma_2 )  \oplus
D^{ladd)}_{SL(3,R)}(1,\sigma_2)$ representation.  However, the
resulting representations obtained  after the $\Gamma^{m}$ action have
different values of the Casimir operators and thus define new
(mutually  orthogonal) Hilbert spaces.

\subsection{Algebraic solution for $\Gamma^{m}$}

A rather efficient way to impose additional algebraic
constraints on the vector operator $\Gamma$ consists in embedding it
into a non-Abelian Lie-algebraic structure. The minimal semi-simple
Lie algebra that contains both the $sl(D,R)$ algebra and the
corresponding vector operator $\Gamma$ is given by the $sl(D+1,R)$
algebra. There are two $SL(D,R)$ vector operators: $A^m$ and $B_m$,
$m=1,2,\dots D$, in the $sl(D+1,R)$ algebra that transform
w.r.t. $[1,0]$ and  $[1,1, \dots ,1]$  representations of $SL(D,R)$
respectively. Components of each of them mutually commute, while their
commutator yields the $SL(D,R)$ generators themselves, i.e.
$$ [A^{m}, A^{n}] = 0,\quad [B_{m}, B_{m}] = 0,\quad [A^{m}, B_{n}] =
i Q^{m}_{n}.
$$ Now, due to the $sl(D+1,R)$ algebra constraints, any irreducible
representation (or an arbitrary combination of them) of $SL(D+1,R)$
defines a Hilbert space that is invariant under the action of an
$SL(D,R)$ vector operator $\Gamma^{m}$ proportional to $A$ or $B$.

\subsection{$\Gamma^{m}$ construction in the $D = 3$ case}

\vskip12pt $\overline{SL}(3,R)$ is embedded into $\overline{SL}(4,R)$,
and a reduction of the spinorial irreducible representations
(multiplicity free Discrete Series) of the latter group down to $D=3$
is as follows [15]:
\begin{eqnarray*}
D^{disc}_{\overline{SL}(4,R)}(j_0,0) &\supset&
\bigoplus_{j=1}^{\infty}
D^{disc}_{\overline{SL}(3,R)}(j_0;\sigma_2(j),\delta_1(j)) \\
D^{disc}_{\overline{SL}(4,R)}(0,j_0) &\supset&
\bigoplus_{j=1}^{\infty}
D^{disc}_{\overline{SL}(3,R)}(j_0;\sigma_2(j),\delta_1(j)) \nonumber
\end{eqnarray*}

The vector operator is either $\Gamma \sim A$ or $\Gamma \sim B$. The
explicit form of the $A + B$ operator (in the spherical basis of the
$Spin(4) = SU(2)\otimes SU(2)$ group) is well known, while the above
embedding approach yields a closed expressions for the $A - B$
operator as well. In particular,
\begin{eqnarray*}
&&\left<\begin{array}{c}J' \\ M'\end{array}\right| (A - B)_{\alpha}
\left|\begin{array}{c}J \\ M\end{array}\right> 
\\
&&\quad\quad = i\sqrt{6}(-)^{J'-M'} \sqrt{(2J'+1)(2J+1)} 
\left(\begin{array}{ccc}\ J'& 1 & J \\ -M' & \alpha & M \end{array}\right) 
\\
&&\quad\quad\quad \times\left\{
\begin{array}{ccc} j'_1 & 1 & j_1 \\ j'_2 & 1 & j_2 \\ J' & 1 & J
\end{array} \right\} <j'_1 j'_2 || Z || j_1 j_2 > ,
\end{eqnarray*}
where, $< j'_1 j'_2 || Z || j_1 j_2 >$ are known reduced matrix
elements of the $\overline{SL}(4,R)$ noncompact operators
$Z_{\alpha\beta}$.

Finally, we can write an $\overline{SL}(3,R)$ covariant spinorial wave
equation in the form
\begin{eqnarray*}
&&(i \Gamma^{m}\partial_{m} - \mu ) \Psi (x) = 0 , \\ && \Psi\ \sim\
D^{disc}_{\overline{SL}(4,R)}(j_0,0),\
D^{disc}_{\overline{SL}(4,R)}(0,j_0) ,  \\ &&\Gamma^{m} =
\frac{1}{2}(J^{(1)m} - J^{(2)m} + (A - B)^{m}) , \quad\quad m = 0, 1, 2
\end{eqnarray*}
The matrix elements of all operators defining the $\overline{SL}(3,R)$
vector operator $\Gamma^{m}$ in the infinite-component representation
of the field $\Psi (x)$ are explicitly constructed.


\begin{thebibliography}{99}

\bibitem{1} E. Bergshoeff, E. Sezgin and P.K. Townsend,
  ``Supermembranes and eleven-dimensional supergravity"
  {\it Phys. Lett. B}  {\bf 189} (1987) 75.

\bibitem{2} M.J. Duff, 
  ``Supermembranes", Lectures given at the Theoretical Advanced Study Institute
  in Elementary Particle Physics (TASI 96), Boulder, 1996, hep-th/9611203.

\bibitem{3} Y. Ne'eman and Dj. \v{S}ija\v{c}ki, 
  ``$\overline{GL}(4,R)$ Group-Topology, Covariance and Curved-Space Spinors'',
  {\it Int. J. Mod. Phys.} {\bf A 2} (1987) 1655.

\bibitem{4} M. Berg, C. DeWitt-Morette, S. Gwo and E. Kramer, 
  ``The Pin Groups in Physics: C, P, and T'',
  {\it Rev. Math. Phys.} {\bf 13} (2001) 953.

\bibitem{5} Dj, \v{S}ija\v{c}ki, ``Affine Particles and Fields'' {\it
  Int. J. Geom. Meth. Mod. Phys.} {\bf 2} (2005) 189.

\bibitem{6} Dj. \v{S}ija\v{c}ki, 
  ``The Unitary Irreducible Representations of $\overline{SL}(3,R)$'', 
  {\it J. Math. Phys.} {\bf 16} (1975) 298.

\bibitem{7} F.W. Hehl, G.D. Kerlick and P. von der Heyde,  
  ``On a New Metric Affine Theory of Gravitation'', 
  {\it Phys. Lett.} {\bf B 63} (1976) 446;
  F.W. Hehl, J.D. McCrea, E.W. Mielke and Y. Ne'eman,
  ``Metric Affine Gauge Theory of Gravity: Field Equations, Noether
  Identities, World Spinors, and Breaking of Dilation Invariance'',
  {\it Phys. Reports} {\bf 258} (1995) 1.

\bibitem{8} Y. Ne'eman and Dj. \v{S}ija\v{c}ki, 
  ``Unified Affine Gauge Theory of Gravity and Strong Interactions with Finite
  and Infinite  $\overline{GL}(4,R)$ Spinor Fields'',
  {\it Ann. Phys. (N.Y.)} {\bf 120} (1979) 292;
  Y. Ne'eman and Dj. \v{S}ija\v{c}ki, 
  ``Gravity from Symmetry Breakdown of a Gauge Affine Theory'',
  {\it Phys. Lett.} {\bf B 200} (1988) 489.

\bibitem{9} Y. Ne'eman and Dj. \v{S}ija\v{c}ki,
  ``Spinors for Superstring in a Generic Curved Space'',
  {\it Phys. Lett.} {\bf B 174} (1986) 165.

\bibitem{10} Y. Ne'eman and Dj. \v{S}ija\v{c}ki,
  ``Curved Space-Time and Supersymmetry Treatments for p-Extendons'',
  {\it Phys. Lett.} {\bf B 206} (1988) 458.

\bibitem{11} V. O. Ogievetskii and I. V. Polubarinov, 
  {\it JETP} {\bf 48} (1965) 1625.

\bibitem{12} Dj. \v{S}ija\v{c}ki,
  ``$\overline{SL}(n,R)$ Spinors for Particles, Gravity and Superstrings'',
  in {\it Spinors in Physics and Geometry}, A. Trautman and G. Furlan eds.
  (World Scientific Pub., 1988) 191;
  Dj, \v{S}ija\v{c}ki, ``Generic Curved Space Superextendon Theories'',
  in {\it Supermembranes and Physics in 2+1 Dimensions}, 
  eds. M. Duff, C. Pope and E. Sezgin (World Scientific Pub., 1990) 213.

\bibitem{13} Y. Ne'eman and Dj. \v{S}ija\v{c}ki, 
  ``$\overline{SL}(4,R)$ World Spinors and Gravity'',
  {\it Phys. Lett.} {\bf B 157} (1985) 275.

\bibitem{14} Dj. \v{S}ija\v{c}ki, 
  ``World Spinors Revisited'',
  {\it Acta Phys. Polonica} {\bf B 29} (1998) 1089.

\bibitem{15} Dj. \v{S}ija\v{cki}, 
  ``$\overline{SL}(4,R)$ Embedding for a 3D World Spinor Equation'', 
  {\it Class. Quant. Grav.} {\bf 21} (2004) 4575.

\bibitem{16} I.Kirsch and Dj. \v{S}ija\v{c}ki, 
  ``From Poincar\'e to Affine Invariance: How does the Dirac Equation
  Generalize?'',
  {\it Class. Quant. Grav.} {\bf 19} (2002) 3157.


\end{thebibliography}
\end{document}